\pdfoutput=1
\documentclass[english]{article}
\usepackage{geometry}
\usepackage{babel}
\usepackage{xcolor}
\usepackage{graphicx}
\usepackage{amsmath,amsfonts}
\usepackage{xcolor}
\usepackage{graphicx,indentfirst,subfigure,epsfig}
 \usepackage{varioref}
 \usepackage{wrapfig}
 \usepackage{subfigure}
 \usepackage{subfigmat}
 \usepackage{amsthm}
\usepackage{hyperref}
\hypersetup{colorlinks, citecolor=red, linkcolor=blue, urlcolor=red, breaklinks=true}
\usepackage{scrextend}
\usepackage[T1]{fontenc}
\usepackage[utf8]{inputenc}
\usepackage{tgtermes}
\usepackage{nomencl}
\usepackage{algorithm}
\usepackage{algorithmic}
\usepackage{booktabs}
\usepackage{longtable}
\usepackage{lineno}
\usepackage{listings}

\definecolor{codegreen}{rgb}{0,0.6,0}
\definecolor{codegray}{rgb}{0.5,0.5,0.5}
\definecolor{codepurple}{rgb}{0.58,0,0.82}
\definecolor{backcolour}{rgb}{0.95,0.95,0.92}

\lstdefinestyle{mystyle}{
    backgroundcolor=\color{backcolour},
    commentstyle=\color{codegreen},
    keywordstyle=\color{magenta},
    numberstyle=\tiny\color{codegray},
    stringstyle=\color{codepurple},
    basicstyle=\ttfamily\footnotesize,
    breakatwhitespace=false,
    breaklines=true,
    postbreak=\mbox{\textcolor{red}{$\hookrightarrow$}\space},
    captionpos=b,
    keepspaces=true,
    numbers=left,
    numbersep=5pt,
    showspaces=false,
    showstringspaces=false,
    showtabs=false,
    tabsize=2
}

\usepackage{lineno,xcolor}
\usepackage[inline]{enumitem}

\usepackage{bm}

\graphicspath{{./}{./figs/}}

  \def\clap#1{\hbox to 0pt{\hss#1\hss}}

\makenomenclature

\begin{document}
\large
\title{\textbf{Jarvis for Aeroengine Analytics: A Speech Enhanced Virtual Reality Demonstrator Based on Mining Knowledge Databases}}

\author{S\l awomir Konrad Tadeja\footnote{Address all correspondence to \texttt{skt40@eng.cam.ac.uk}}$\;^{\star \dagger}$, 
Krzysztof Kutt$^{\dagger}$, 
Yupu Lu$^{\star}$, 
Pranay Seshadri$^{\ddagger \smallint}$, \\
Grzegorz J. Nalepa$^{\dagger}$,
Per Ola Kristensson$^{\star}$\\ 
\vspace{0.3 cm}\\ 
$^{\star}$Department of Engineering, University of Cambridge, Cambridge, U. K., \\ 
$^{\dagger}$Institute of Applied Computer Science, Jagiellonian University in Krak\'{o}w, Krak\'{o}w, Poland, \\
$^{\ddagger}$Data-Centric Engineering, The Alan Turing Institute, London, U. K. \\
$^{\smallint}$Department of Mathematics, Imperial College, London, U. K.}
\date{}
\maketitle{}

\begin{abstract}
In this paper, we present a Virtual Reality (VR) based environment where the engineer interacts with incoming data from a fleet of aeroengines. This data takes the form of 3D computer-aided design (CAD) engine models coupled with characteristic plots for the subsystems of each engine. Both the plots and models can be interacted with and manipulated using speech or gestural input. The characteristic data is ported to a knowledge-based system underpinned by a knowledge-graph storing complex domain knowledge. This permits the system to respond to queries about the current state and health of each aeroengine asset. Responses to these questions require some degree of analysis, which is handled by a semantic knowledge representation layer managing information on aeroengine subsystems. This paper represents a significant step forward for aeroengine analysis in a bespoke VR environment and brings us a step closer to a Jarvis-like system for aeroengine analytics.
\end{abstract}

\section{INTRODUCTION}
Our work presented in this paper has been inspired by the popular culture, specifically by well-known science-fiction novels and movie franchises. In these movies, the protagonists, are seen visualizing data, drawing inference and making decisions based on interactions with the some sort of artificial intelligence (AI) assistant, such as Jarvis. The ability to query such an assistant---particularly for quantitative insights---is appealing and can potentially reduce engineering-hours with the promise of more efficient and analytically-grounded decisions.

\begin{figure}[!th]
    \centering
    \includegraphics[width=\textwidth]{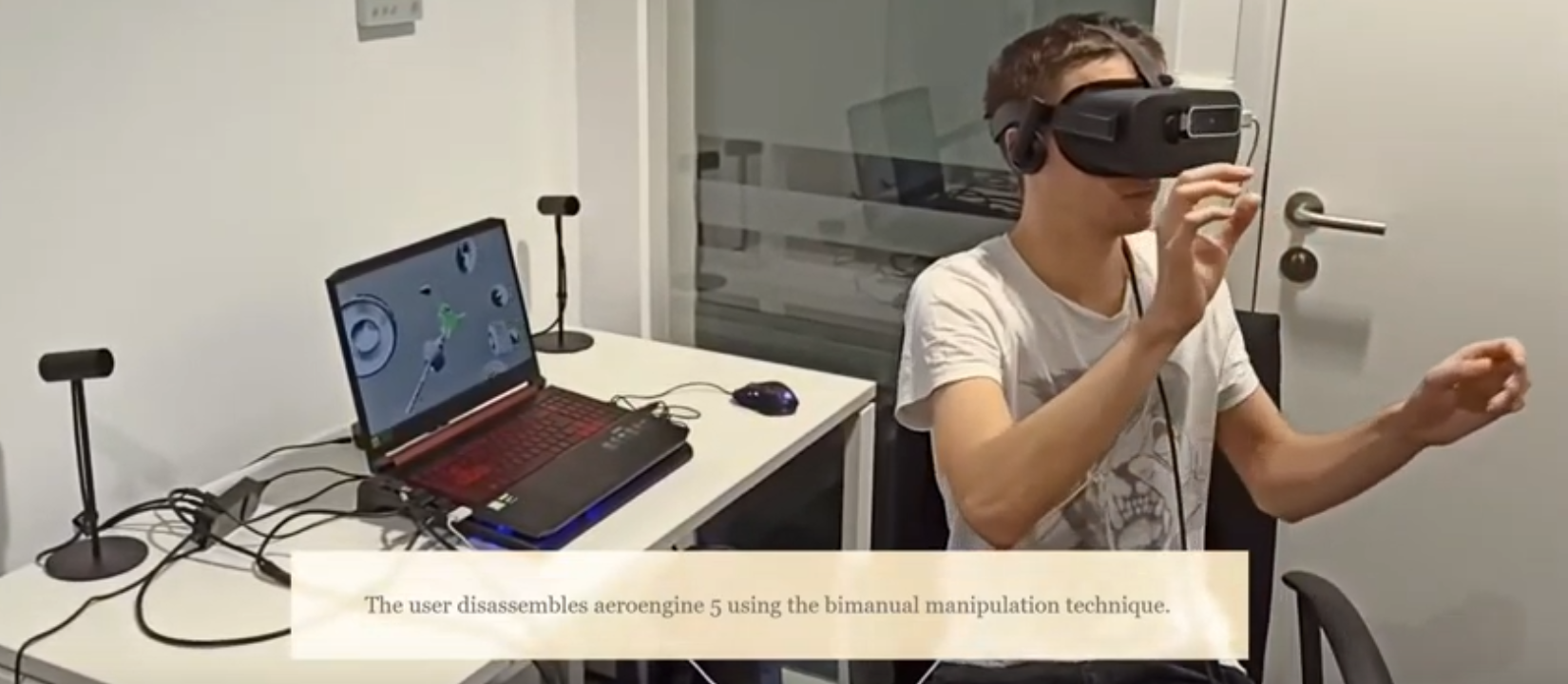}
    \caption{A video snapshot showing the user operating the Jarvis system \cite{video_snap}. The user's VR view is shown on the laptop screen. The VR HMD is an Oculus Rift including two Oculus sensors. Hand-tracking and gesture recognition is facilitated by a Leap Motion sensor attached to the front of the HMD. Voice interaction uses Rift's built-in speakers and microphone.}
    \label{fig:video}
\end{figure}

In this paper we develop a bespoke 3D interface-based~\cite{sutherland_head-mounted_1968} immersive virtual reality environment, where the engineer interacts with incoming data from a fleet of aeroengines~\cite{tadeja_ieee_2020}. This data is ported to a knowledge-based system underpinned by a knowledge graph that stores complex turbomachinery-specific domain knowledge. This system is capable of responding to queries such as, ``What is the pressure ratio of the intermediate pressure compressor for Engine A{?}'' or, ``Where is the engine with the highest average efficiency of fleet B currently{?},'' or, ``Compute the average pressure ratio after 100 hours of flying time for component C in Engine D''. Responses to these questions not only involve access to data but also require some degree of analysis. Our knowledge-based system is able to handle such queries. Further, a degree of flexibility is embedded that permits the user to construct queries from a broader lexicon. The key idea is that we have leverage over more contemporary deep-learning based systems, because our subject matter is very specific and the number of topics limited.

The knowledge-based system described here provides a semantic knowledge representation layer for capturing information. This knowledge is used for dynamic hinting during design and visualization. The representation mechanism uses a triple-set representation (subject-predicate-object) using the RDF (Resource Description Framework) language~\cite{rdf11-primer}. A set of triples creates a (directed) knowledge graph, where subject and object are nodes and predicate describes an edge. In such a graph, automatic reasoning is used to infer new knowledge that is derived from existing knowledge, or to verify the aptness of the knowledge base. Knowledge in the knowledge graph can be accessed using an SQL-like language called SPARQL~\cite{sparql11-overview}. Voice-to-SPARQL translation is a very complex task. However, owing to the fact that the system has a very limited scope, we can provide an interface based on a set of interaction templates. This approach is more reliable than a generic voice recognition system while still allowing a natural way of interaction from the user's point of view. There have been some prior attempts to utilize voice-operated interfaces, or particular parts of it, such as speech recognition, natural language processing (NLP) or speech synthesis in the aerospace domain, see, for instance, \cite{883_1988, 332846_1994, 821658_2000, 7492684_2015, 7778024_2016}.

This paper represents one of the first forays of aeroengine analysis in a bespoke VR environment and brings us a step closer to a voice-operated assistant for aeroengine analytics. A description of an earlier, non-voice operated version of this system is provided in Tadeja et al.~\cite{tadeja_ieee_2020, tadeja_aerovr}. The snapshot from the  video-clip\footnote{\url{https://youtu.be/TmjmRD8YJYI}} showcasing our work, as seen in Fig.~\ref{fig:video}, is available online \cite{video_snap}.

\section{VIRTUAL REALITY FOR AERONAUTICS AND ASTRONAUTICS}

Some of the earliest examples of how to exploit 3D interfaces within the aerospace domain can be traced back to Hale~\cite{hale_applied_1994} and Mizell~\cite{mizell_virtual_1994}. In both cases, the authors lay the groundwork for research concerning the utility of applying VR in design.

Garc\'ia-Hernandez et al.~\cite{garcia-hernandez_perspectives_2016} identify three related areas in which the usage of VR holds the most promise: (i) the extraction of information from coupling multiple 2D data into 3D-like structures; (ii) visualization of perplexed, non-planar graphs (for a survey of graph visualization techniques see Herman et al.~\cite{841119}); and (iii) the 3D version of the parallel coordinates plots (PCP)~\cite{146402}. Examples of VR-based \textit{immersive parallel coordinates plots} (IPCP) used to visualize datasets from aerodynamic design case studies can be found in Tadeja et al.~\cite{tadeja_chi_2019} and Tadeja et al.~\cite{tadeja_aiaa_2020}. Further, Garc\'ia-Hernandez et al.~\cite{garcia-hernandez_perspectives_2016} recognize a number of areas of VR which currently are, or have the potential, to be successfully exploited, including spacecraft optimal path analysis~\cite{964562}; exoplanet research~\cite{refId0}; mission planning; launch/abort decision making~\cite{10.1145/1014052.1014104}, space debris, asteroid and near earth object (NEO) orbit analysis~\cite{Meador}; citizen science~\cite{10.1525/bio.2009.59.11.9} in aerospace; and spacecraft design optimization~\cite{stump_trade_2004} as well as aerodynamic design~\cite{jeong_data_nodate}. Other, not listed, include collaborative environments~\cite{roberts_collaborative_2015, clergeaud_3d_2016}; haptic systems~\cite{savall_description_2002, sagardia_interactive_2013}; aerospace simulation~\cite{stone_evolution_2011}; planetary exploration~\cite{wright_immersive_2001}; and management of aerospace telemetry and sensor data  \cite{wright_immersive_2001,russell_acquisition_2009,lecakes_visualization_2009}.

In particular, the latter application area is closely related to what we are trying to achieve with the help of our immersive system for aeroengine analytics in VR. For instance, Wright et al.~\cite{wright_immersive_2001} propose the idea of interlinking CAD models with sensor and telemetry data where a change in such a data stream would lead to an appropriate change in the CAD visualization as well. Both Russell et al.~\cite{russell_acquisition_2009} and Lecakes et al.~\cite{lecakes_visualization_2009} present approaches for using VR environments for system diagnosis and health management for rocket engine tests.

\section{SYSTEM ARCHITECTURE}
\begin{figure}[!ht]
    \centering
    \includegraphics[width=\textwidth]{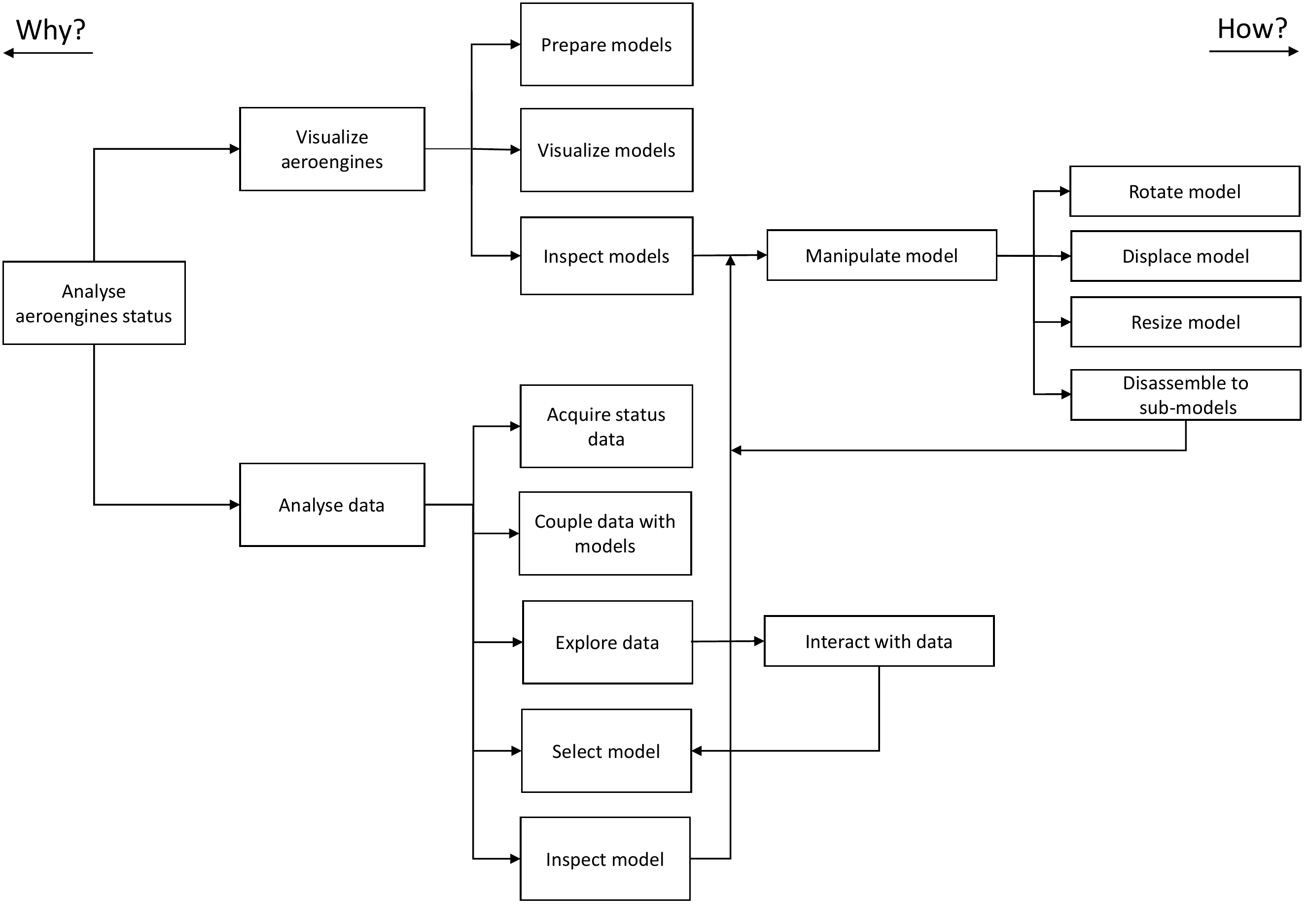}
    \caption{The FAST (\textit{Function Analysis Systems Technique})~\cite{Shefelbine_edc} diagram of the Jarvis system. Diagram adapted from Tadeja et al.~\cite{tadeja_aerovr} with permission.}
    \label{fig:fast}
\end{figure}

In this section we present and discuss the overall system architecture of the Jarvis system by decomposing a function model using using FAST (Function Analysis Systems Technique)~\cite{Shefelbine_edc} and by analyzing signals between the main function structures. We also present a brief task analysis to distill the main tasks of the user interacting with the system.

\subsection{Function Model}

Fig.~\ref{fig:fast} shows a FAST-diagram decomposing the main functions of the Jarvis system. The FAST-diagram should be read from left-to-right going top-to-bottom. We use the FAST analysis to carry out a brief task analysis later in this section. The FAST-analysis is carried out at the functional level to avoid design fixation and does not make any explicit reference to solutions (function carriers).

\subsection{System Model}
\begin{figure}[!ht]
    \centering
    \includegraphics[width=\textwidth]{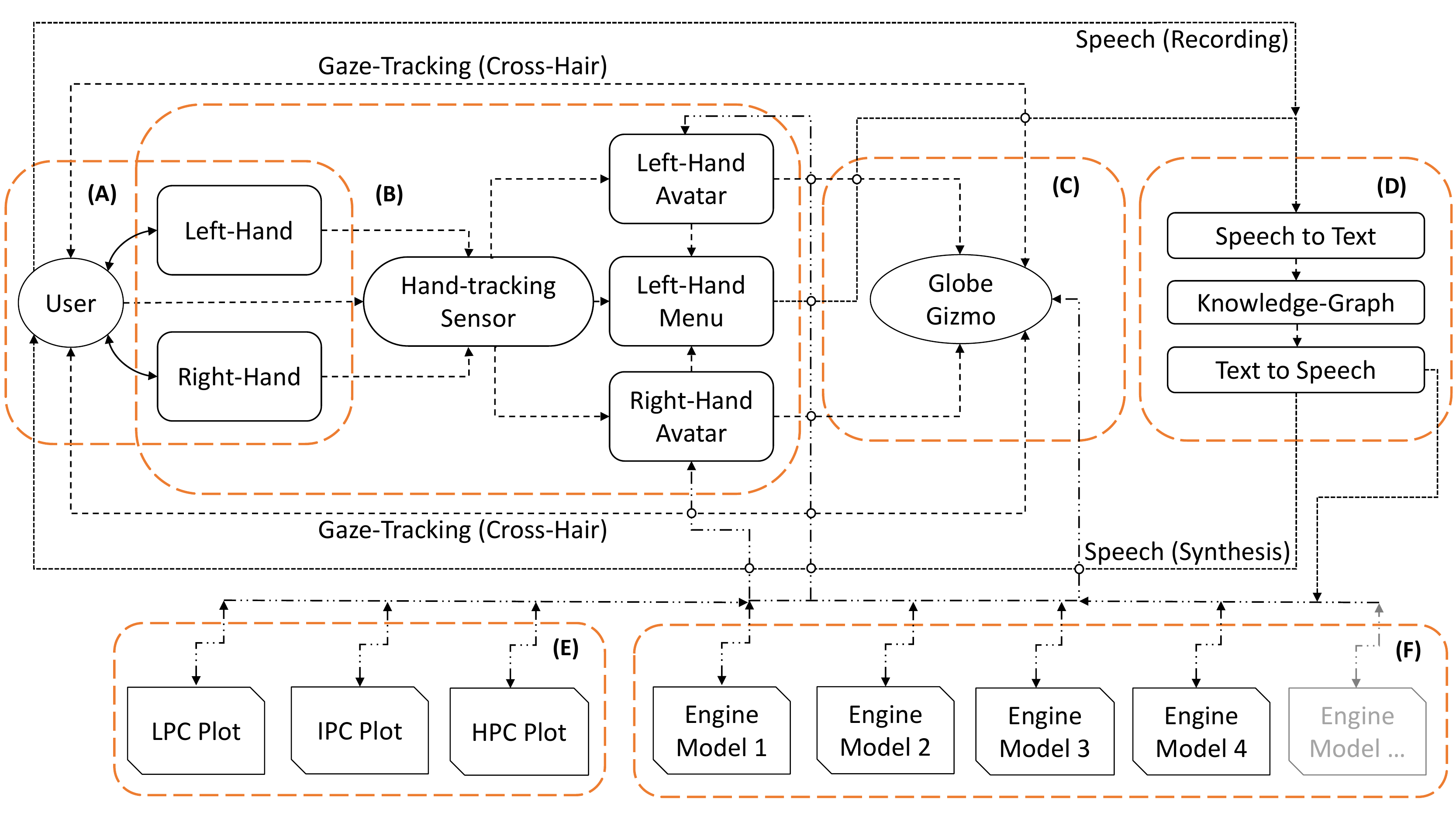}
    \caption{The diagram of the signals flow between the five key components of the Jarvis system. These elements are: (a) the user; (b) the hand-tracking bundle; (c) the globe gizmo; (d) the speech bundle; and (e) the engine 3D models. The signal can have a form of visual clues (e.g.~cross-hair, hand avatars) and stimuli (e.g.~object's highlights), audio (recorded and synthesized speech) and others. The diagram is adapted from Tadeja et al.~\cite{tadeja_isprs_2019} with permission.}
    \label{fig:signals}
\end{figure}

We study our Jarvis system model by analyzing signal flow. Here we pinpoint, list, analyze, and discuss all the uni- and bi-directional signals sent and received by the key components of our system. The signals flowing between various VR elements are shown in Fig.~\ref{fig:signals}. The six key components are the following:

\begin{enumerate}[label=(\Alph*),leftmargin=2.5\parindent] 
\item\label{user_sig} \textbf{The user:} The user interacts with the system using a range of paired methods. These are: (1) gaze-tracing \& ray-tracing; (2) hand-tracking \& gesture-recognition; and finally, (3) speech-recognition \& speech-synthesis. All these pairs exhibit bi-directional signal flow as shown in Fig.~\ref{fig:signals}. To approximate gaze-tracking we use a cross-hair placed at a fixed distance from the user's gaze (see Fig.~\ref{fig:full_view}). From the center of the cross-hair we continuously extend rays invisible to the user. Whenever the cross-hair circles over an interactive object, the system recognizes this situation and the object is highlighted---indicating to the user that it can be interacted with. Direct manipulation of the objects is achievable with the user's hands. The gestural input is facilitated with hand-tracking that communicates in real-time with the virtual hand representations, which are in turn signaling the hand's current position and gesticulation to the user (see Fig.~\ref{fig:gestures}). This mechanism is discussed in more detail in the description of the hand-tracking component (item B). Finally,  speech-recognition and speech-synthesis are supported via a microphone and speakers integrated in the headset and a range of external resources. This mechanism will also be described in more detail in the description of the speech component (item D).

\item\label{hands_sig} \textbf{The hand-tracking bundle:} To interact with the Jarvis system the user is required to maintain their hands within the Leap Motion~\cite{leapmotion} hand-tracking sensor's field of view. This tracking state is signaled to the user using the virtual representation of the user's hands---the hands avatars (see Fig.~\ref{fig:globe_gizmo} and Fig.~\ref{fig:compressors}(a)). The hand avatars with the visualized joints will instantaneously follow and emulate the user's hand gestures, hence, the hand-tracking bundle is in a constant signal loop with the user. An articulated gesture will immediately influence the currently selected interactive object, allowing the user to release the handle over an object, or select and manipulate another object. Object selection and deselection is signaled to the user by manipulating the highlight state of objects.

\item\label{gizmo_sig} \textbf{The globe gizmo:} The globe gizmo consist of three components: (1) the simplified globe model~\cite{globe_asset}; (2) a number of interactive 3D models of the airplanes~\cite{airplanes_asset} circling around the globe on predefined trajectories; and (3) the movement selector for the gizmo (a red sphere placed in a fixed position above the gizmo). All these elements can be seen in Fig.~\ref{fig:globe_gizmo}. Both the movement selector and the airplane models are interactive and respond to the user's gaze tracking. Airplane models can be selected and highlighted in response to user actions over \ref{plots_sig} the compressors characteristics, \ref{engine_sig} the engine 3D models, or in response to \ref{speech_sig} the user's voice commands.

\item\label{speech_sig} \textbf{The speech bundle:} The user interacts with the speech-bundle by pressing a button on the left-hand pop-up menu (see Fig.~\ref{fig:gestures}). Next, once the user has formulated a query and spoken it, the acoustic signal is decoded into text and fed to the the knowledge graph. The outcome of a query processed within the knowledge-graph is signaled to the user using a text-to-speech service. If appropriate, this outcome may also be signaled using an automatic selection of an appropriate \ref{engine_sig} aeroengine model, as well as \ref{gizmo_sig} an airplane model on the globe gizmo, and/or \ref{plots_sig} a marker on one of the compressors characteristics. Further, if the speech-to-text translation fails this event is also signaled to the user with a voice message.

\item\label{plots_sig} \textbf{The compressors characteristics plots:} The compressor characteristics plots are constructed using three elements: (1) interactive movement selectors in the form of large spherical markers in the top-right corner; (2) color-coded static contour lines representing constant speed lines; and (3) interactive small spherical markers representing operating points for each engine. All these elements can be seen in Fig.~\ref{fig:full_view}.

\item\label{engine_sig} \textbf{The engine 3D models:} The aeroengine 3D CAD models~\cite{heyns} (see Subsection~\ref{AeroModels}) are highlighted in response to the user's gaze to signal to the user that they can be interacted with. Once selected, they switch their color for a couple of seconds to inform the user that they have been selected (see Fig.~\ref{fig:compressors}). Such signaling can also take place in the system's response to a \ref{speech_sig} voice command issued by the user after \ref{hands_sig} pressing the appropriate button on the left-hand pop-up menu (see Fig.~\ref{fig:gestures}(a)) or upon selection through the \ref{gizmo_sig} globe gizmo. In response to \ref{hands_sig} hand-tracking, bi-manual manipulation carried out by the user over an engine model which results in in a change in either a model's position, shape, or rotation \cite{tadeja_ieee_2020} is immediately visible.
\end{enumerate}

\subsection{Task Analysis}\label{tasks}
Using the FAST-diagram as a base, we distill two key user tasks:

\textbf{T1---Visualize aeroengines}: The system should simultaneously visualize an array of aeroengine 3D models (see Subsection~\ref{AeroModels}) together with their accompanying data.

\textbf{T2---Analyze data}: The system should support and aid the user in effective and efficient analysis of the performance data associated with the given components of the aeroengines. Here we focus on the LPC, IPC, and HPC components (see Subsection~\ref{AeroModels}).

\noindent These two tasks---\textit{T1} and \textit{T2}---can be split into a series of sub-tasks as the data pertaining to each aeroengine is coupled with the 3D models of the individuals engines (see Subsection~\ref{AeroModels}). For instance, in the case of one particular aeroengine, its IPC sub-component has been damaged, which is also reflected in the associated graph. To see such connections the user has to inspect both the data and the models in either order. With this in mind, we list the following additional sub-tasks:

\textbf{T3---Explore performance data}: As each of the aeroengine 3D models is associated with performance characteristics of the selected engine parts (see Subsection~\ref{CompressorsPlots}) the system should support the user in an effective exploration of such data. Here we are focusing on the LPC, IPC, and HPC parts (see Subsection~\ref{AeroModels}) coupled with the compressors characteristics plots (see Subsection~~\ref{CompressorsPlots}).

\textbf{T4---Select aeroengine model}: The system should allow the user to easily select the desired aeroengine from an array of CAD aeroengines 3D models (see Subsection~\ref{AeroModels}).

\textbf{T5---Inspect aeroengine model}: The system should support the user in efficient inspection of a given aeroengine 3D model (see Subsection~\ref{AeroModels}).

\textbf{T6---Manipulate aeroengine model}: The system should support a form of manipulation of the individual CAD models. This, in turn, should aid the user in fulfilling the tasks \textit{T3}, \textit{T4}, and \textit{T5}.

\section{APPARATUS \& VISUALIZATION FRAMEWORK}

\subsection{System Structure}
\begin{figure}[!ht]
    \centering
    \includegraphics[width=\textwidth]{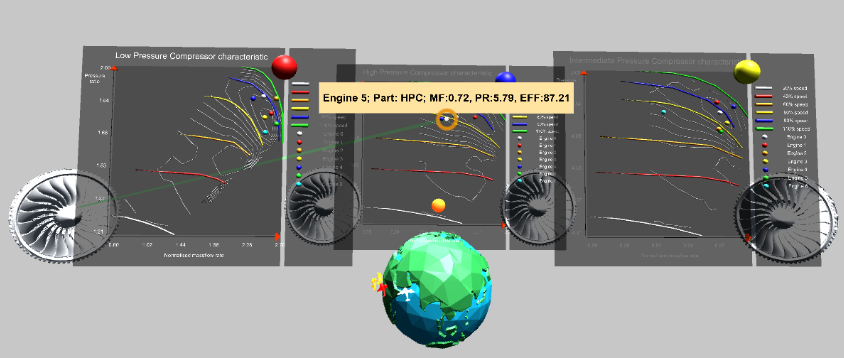}
    \caption{An overview of the AeroVR system as seen by the user from the initial viewing position. The four aeroengines are visible in the back, coupled with the compressors characteristics plots (LPC, IPC and HPC from left to right respectively) and the globe gizmo~\cite{globe_asset} in the middle. The user's gaze is focused on a marker on the HPC plot via an orange cross-hair. The pop-up textbox informs the user about the operating point of this particular HPC on engine 5 and simultaneously draws a line (in the same color as the marker, that is, light green) pointing towards the appropriate turboengine's CAD 3D model~\cite{heyns}.}
    \label{fig:full_view}
\end{figure}

The visualization framework used to support the Jarvis system was first developed to explore and assess the feasibility of using digital twins of aeroengines embedded in a VR environment~\cite{tadeja_ieee_2020}. Details of the development and verification of this system, as well as the results of a qualitative user study with domain experts, can be found in Tadeja et al.~\cite{tadeja_ieee_2020}. Here, we discuss how this framework has been extended and updated with additional features. Those extensions allow the user to interact with the objects and the system itself in a novel voice-based manner. We have also switched the initial point of the user's focus to a globe-like interactive gizmo that allows the user to track the trajectory of the individual airplanes in real-time (see Fig.~\ref{fig:globe_gizmo}).

\subsection{Hardware}
The Jarvis VR system was developed and tested on a computer setup consisting of NVIDIA GeForce GTX 1060 GPU coupled with the Oculus Rift head-mounted display (HMD) used to facilitate the VR environment. The hand-tracking was achieved using a Leap Motion sensor~\cite{leapmotion} attached to the front of the headset.

\subsection{Visualization}
\begin{figure}[!htp]
    \centering
    \includegraphics[width=\textwidth]{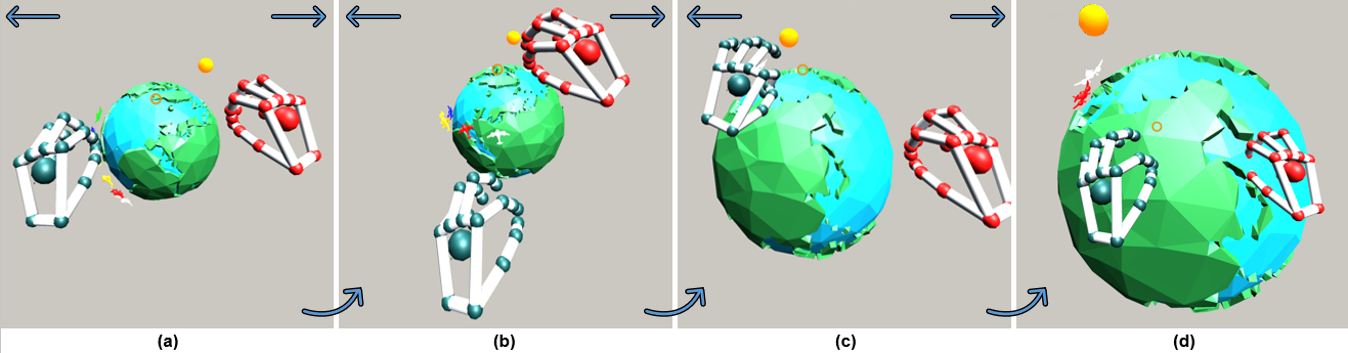}
    \caption{The interactive globe gizmo~\cite{globe_asset} with airplane models~\cite{airplanes_asset} circling it on a given predefined trajectory. If coupled with real-life streamed data, such as the flight paths, current positions and speed, it could be used to provide the user visual aid in tracking the current status of a fleet of airplanes. The user can select the gizmo by gazing over the movement selector (an orange marker positioned above the North Pole) and making a double-pinch gesture~\cite{tadeja_ieee_2020}. The user can bimanually manipulate the model (a)---(d). For example, the user can rotatea model, decrease or increase its size, or move it to a different position in 3D space. Airplane models~\cite{airplanes_asset} can be selected in the same way as the gizmo and will upon selection automatically highlight and select the appropriate aeroengine CAD model~\cite{heyns}.}
    \label{fig:globe_gizmo}
\end{figure}

The visualization was developed using the Unity  engine~\cite{unity} and is based on the VR system for digital twinning, details of which can be found in Tadeja et al.~\cite{tadeja_ieee_2020}. This system was updated and extended with a range of new features and visualization components. It now includes a speech-based interface and the globe-based interactive gizmo (see Fig.~\ref{fig:globe_gizmo}) that can be used to track the current geoposition of a fleet of aeroengines around the world.

\subsubsection{Aeroengines Models}\label{AeroModels}
\begin{figure}[!htp]
    \centering
    \includegraphics[width=\textwidth]{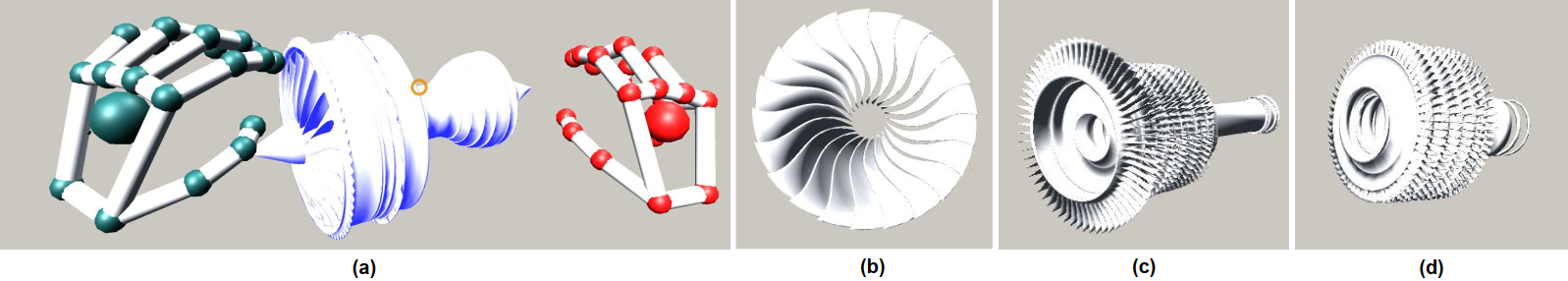}
    \caption{The aeroengine 3D CAD model~\cite{heyns}: (a) a full aeronengine model in a selected state with hand avatars manipulating the model; (b--d) low (LPC), intermediate (IPC), and high pressure (HPC) compressors. For clarity, the respective sizes of the compressors with regards to each other, as well as to the engine model shown in (a), are not preserved in this figure.}
    \label{fig:compressors}
\end{figure}

Each of the aeroengine CAD models~\cite{heyns} can be disassembled into these eleven, non-separable sub-parts:
\begin{enumerate*}[label=(\alph*)]
    \item casing;
    \item low pressure compressor (LPC);
    \item intermediate pressure compressor (IPC);
    \item high pressure compressor (HPC);
    \item low pressure turbine, shaft and nozzle;
    \item fan;
    \item nose cone;
    \item high pressure shaft;
    \item intermediate pressure shaft;
    \item intermediate pressure turbine; and
    \item combustor and high pressure turbine.
\end{enumerate*}

The key sub-components of each model are the low pressure compressor (LPC) (see Fig.~\ref{fig:compressors}(b)), the intermediate pressure compressor (IPC) (see Fig.~\ref{fig:compressors}(c)), and the high pressure compressor (HPC) (see Fig.~\ref{fig:compressors}(d)).

\subsubsection{Compressors Characteristics Plots}\label{CompressorsPlots}
Compressor characteristics are standard plots within the turbomachinery community for assessing the performance of a compressor. The horizontal axis of these plots typically shows a non-dimensionalized massflow rate (or flow function), while the vertical axis shows the pressure ratio. Data is typically shown at multiple speed lines, corresponding to different compressor shaft speeds. Iso-contours of efficiency are also plotted, which, in conjunction with the pressure ratio and non-dimensionalized massflow rate, offer a global perspective of a compressor's performance.

The compressors characteristics plots can be seen in Fig.~\ref{fig:full_view}. Further details about those plots as well as a description on how they can be manipulated using hand-tracking \cite{leapmotion} can be found in Tadeja et al.~\cite{tadeja_ieee_2020}.

\section{MULTIMODAL INTERFACE}
The system combines three interaction modalities: 1) gaze-tracking; 2) hand-tracked gestures; and 3) speech recognition.

\begin{figure}[!htp]
    \centering
    \includegraphics[width=\textwidth]{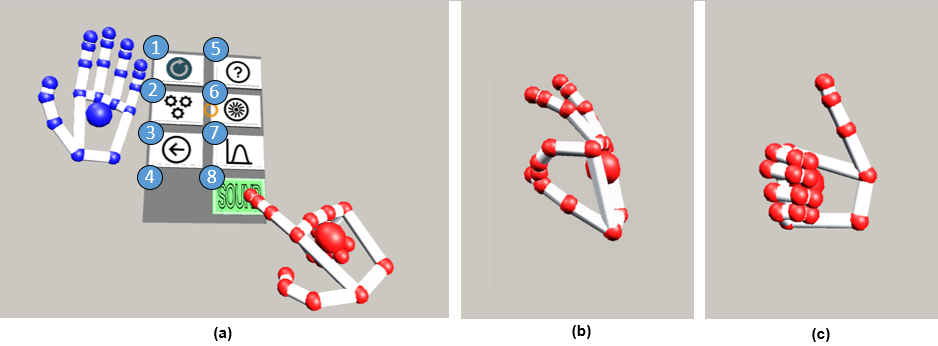}
    \caption{The gestures recognized by the system: (a) the left-hand pop-up gesture and the button press gesture; (b) the single or double pinch gesture; and (c) the thumb-up gesture. For the left-hand menu shown in (a) the menu options are as follows: (1) [$RESET$]---resets the entire visualization to its default state; (2) [$GEAR\_MODE$]---switches to a system mode in which the user can pull an engine model apart; (3) [$REVERT$]---is used to automatically assemble the previously dissembled engine model; (4) [$EMPTY$]---an empty placeholder that allows another button to be placed there; (5) [$HELP$]---displays the help menu; (6) [$ENGINE\_MODE$]---switches to a system mode where the user can manipulate an engine model as a composite unit; (7) [$PLOTS$]---toggles between visible and invisible compressors characteristics plots; and (8) [$SOUND$]---instructs the Jarvis system to record a voice command and respond to the user's spoken query. Subfigure (b) shows the gesture for selecting and manipulating an interactive element of the visualization using either one hand or both hands. Subfigure (c) shows the gesture for releasing a hold on a selected object. Icons by \textit{Icons8}~(\protect\url{https://icons8.com}).}
    \label{fig:gestures}
\end{figure}

\subsection{Head- and Gaze-Tracking}
Head-tracking is provided by the Oculus Rift HMD \cite{oculus}. As Oculus does not support built-in gaze-tracking capabilities, we approximate the user's current gaze location by placing a cross-hair in the center of the user's field of view at a fixed distance from the user (see Fig.~\ref{fig:full_view}). Whenever the cross-hair passes over an interactive object, the object is automatically highlighted to signal its interactive status. This part of the system is based on the Unity VR Free Sample Pack~\cite{unity_vr_samples_pack}.

\subsection{Hand-Tracking and Gesture Recognition}
The hand-tracking and gesture recognition processing is achieved with the help of the Leap Motion sensor and its software development kit (SDK)~\cite{leapmotion}. The range of gestures recognized by our system is reported in Tadeja et al.~\cite{tadeja_ieee_2020} and consists of (i) a pinch gesture; (ii) a double-pinch gesture; (iii) a left-hand palm-up gesture; and (iv) a thumbs-up gesture. These gestures enable the user to: (a) pull the engine apart using their hands; (b) decrease or increase the sizes of the whole engine or its individual sub-components; and (c) rotate the whole engine or any of its sub-components~\cite{tadeja_ieee_2020}. The same manipulation techniques can be applied to the compressors characteristics plots (see Fig.~\ref{fig:full_view}) as well as the globe gizmo (see Fig.~\ref{fig:globe_gizmo}). However, these two elements cannot be pulled apart. An example of a user manipulating and disassembling an aeroengine model can be seen in Fig.~\ref{fig:compressors}. The details of this bimanual manipulation techniques of the 3D models is available in Tadeja et al.~\cite{tadeja_ieee_2020}.

\subsection{Voice Recognition and Information Capture}\label{voice}
The voice interface is based on the observation that our subject matter is very specific and the number of topics that can be mentioned while the user interacts with the system is limited. There is therefore no need to use a very large vocabulary speech recognition system, which would raise many challenges related to grammatical decomposition, understanding of concepts and their relation to the knowledge base, etc. Instead, we provide a set of specific queries that can be used to interact with the system. During the operation of Jarvis we only match the user's voice query with prepared patterns that are stored in the system.

Having first used a voice-to-text service to decode the user's spoken query to text, we use then use the Rasa NLU (Natural Language Understanding) module of Rasa~\cite{bocklisch2017rasa}, an open source set of Python libraries for building human-machine conversational interfaces. Rasa is based on machine learning models. Therefore, in the beginning, we prepared an appropriate sample of possible user queries from  potential users, which was used to bootstrap the system by training an appropriate model. As can be seen in the excerpt (see Listing~\ref{lst:rasanlu}), the sample file consists of several groups starting with \lstinline{## intent:[NAME]}. Each of these groups represents different types of user queries. We provide a list of sample sentences for each intent. There is a possibility to put placeholders in the queries in the form \lstinline{[VALUE](NAME)}, where \lstinline{NAME} is the name of the placeholder and \lstinline{VALUE} is one of the possible values. Training the model on such a file is only done once. It is carried out offline before the system is used for the first time. Thereafter the trained model is used to parse a given sentence and extract information about the intent name and the identified placeholder's values is returned by Rasa NLU module.

\begin{lstlisting}[caption=An excerpt from RASA NLU training file., label=lst:rasanlu, float=ht, basicstyle=\footnotesize\ttfamily]
## intent:show_engine
- Show engine [0](engine_name).
- Show engine [1](engine_name).
- Show me engine [2](engine_name).
- Show me engine [3](engine_name).
- Where is the [fourth](engine_name) engine right now?
- Where is the [fifth](engine_name) engine right now?

## intent:get_engine
- Identify which engine's [fan](subsystem) is operating at [74](num_value)% [shaft speed](characteristic).
- Identify which engine's [LPC](subsystem) is operating at [104](num_value)% [speed](characteristic).
- Identify which engine's [IPC](subsystem) is operating at [one hundred](num_value) [efficiency](characteristic).
- Identify which engine's [HPC](subsystem) is operating at [99](num_value) [efficiency](characteristic).

## intent:get_value
- At roughly what [speed](characteristic) is engine [1](engine_name)'s [LPC](subsystem) running at?
- At roughly what [speed](characteristic) is engine [7](engine_name)'s [IPC](subsystem) running at?

## intent:closest
- Identify which engine's [IPC](subsystem) is the closest to [choke](subsystem_state).
- Identify which engine's [HPC](subsystem) is the closest to [stall](subsystem_state).
- Identify which engine's [fan](subsystem) is operating dangerously close to [choke](subsystem_state).
- Identify which engine's [LPC](subsystem) is operating dangerously close to [stall](subsystem_state).

## intent:the_best
- Which engine's [fan](subsystem) is running at the [lowest](best_direction) [efficiency](characteristic)?
- Which engine's [LPC](subsystem) is running at the [highest](best_direction) [efficiency](characteristic)?
- Which engine's [IPC](subsystem) is running at the [lowest](best_direction) [efficiency](characteristic)?
- Which engine's [HPC](subsystem) is running at the [highest](best_direction) [efficiency](characteristic)?
\end{lstlisting}

The voice interface workflow is as follows:

\begin{enumerate}
    \item The user selects the ``speak'' button in the VR interface. This is done only for the prototype. In the final version, the microphone will be constantly monitored and commands will be recognized on-the-fly.
    \item The audio signal is captured using a microphone.
    \item The issued voice command is decoded into text using the  \verb|SpeechRecognition|\footnote{\url{https://pypi.org/project/SpeechRecognition}} Python library coupled with the speech-to-text service offered by the Google Cloud Speech API.
    \item The recognized user command is transmitted the knowledge graph server over simple JSON-RPC interface over HTTP. The entire speech bundle (voice recognition and the knowledge graph) is implemented as a separate HTTP server to facilitate remote work of multiple users in the system. The entire knowledge base is therefore located in one place accessible via, for example, a local network.
    \item The speech bundle server processes the command (using RASA NLU, as described above, and the knowledge graph, as described in Sect.~\ref{sec:knowgraph}) and prepares a system response.
    Sample commands sent to the speech bundle server and system responses (based on the knowledge graph prepared for the demo use case) are presented in Listing~\ref{lst:communication}.
    \item The system response is transmitted via a JSON-RPC interface. It consists of three elements: \lstinline{engineID} and \lstinline{subsystem} indicate which part of the visualization should be presented to the user, while \lstinline{message} is communicated to the user using a text-to-speech service. Text-to-speech is performed using the eSpeak Speech Synthesizer\footnote{\url{http://espeak.sourceforge.net/}} used under version 3 of the GNU General Public License.
\end{enumerate}

\begin{lstlisting}[caption=Sample commands sent to speech bundle server and answers returned by the server., label=lst:communication, float=htp]
Command: Which engine's HPC is running at the highest efficiency?
Answer: { "engineID": 0,
          "subsystem": "HPC",
          "message": "HPC of engine 0 has the highest value of Efficiency. It is equal to 88.1635" }

Command: At what speed is HPC of Engine 3 running at?
Answer: { "engineID": 3,
          "subsystem": "HPC",
          "message": "HPC of engine 3 has Speed equal to 80.0" }
\end{lstlisting}

\section{Knowledge Graph}
\label{sec:knowgraph}
Domain knowledge, such as the characteristics of the aeroengine models, is stored in a knowledge-based system. We used the RDF (Resource Description Framework) language~\cite{rdf11-primer} as the knowledge representation mechanism. We chose this approach as it is a standardized formalism that is easy to link with other knowledge bases~\cite{hogan2020knowledge}, which can be useful for further project development. For example, for integration with an external engine's specifications. In RDF each statement is represented as the \verb|Subject-Predicate-Object| triplet (for sample RDF triples, see Tab.~\ref{tab:triples}). We use IRIs (Internationalized Resource Identifiers) to allow for unambiguous identification of resources. A triplet's \verb|Object| can be also be represented as a literal, such as \verb|string|, \verb|number| or \verb|date|.

\begin{table}[htp]
    \centering
    \begin{tabular}{c|c|c}                                \hline \hline
        \emph{Subject} & \emph{Predicate} & \emph{Object} \\ \hline
        Subsystem      & isPartOf         & Engine        \\
        IPC            & isSubclassOf     & Compressor    \\
        Compressor     & isDepictedOn     & Plot          \\ \hline
        Compressor     & PressureRatio    & double        \\
        Compressor     & Speed            & double        \\
        Engine         & VR\_ID           & integer       \\ \hline \hline
    \end{tabular}
    \caption{Sample RDF triples for the aeroengine domain. The first three triplets are object properties---the object is an IRI. The last three triplets are data properties---the object is a literal.}
    \label{tab:triples}
\end{table}

\begin{figure}[htp]
    \centering
    \includegraphics[width=\textwidth]{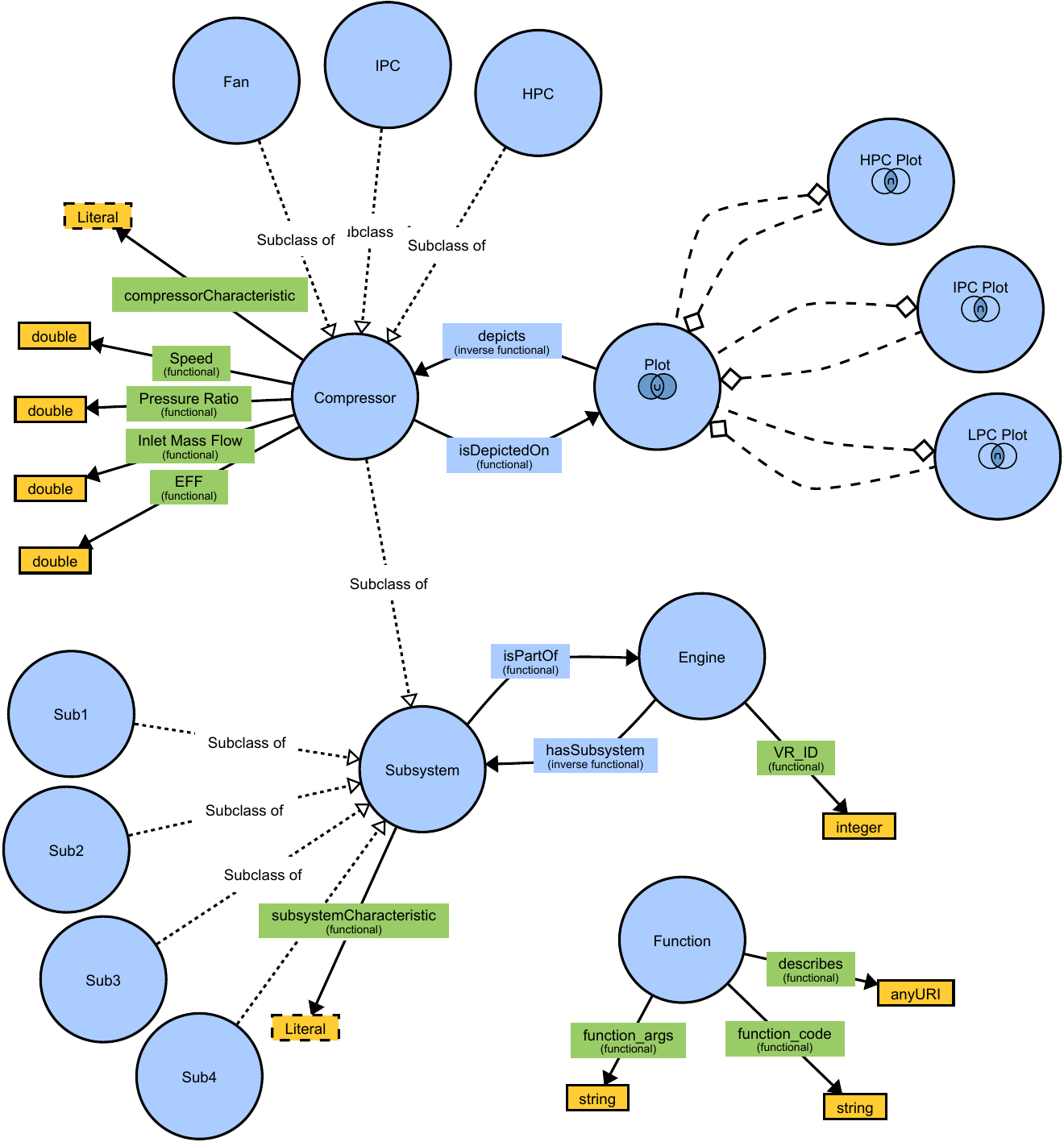}
    \caption{TBox with main concepts and relations used in presented system.}
    \label{fig:graph}
\end{figure}

A set of triplets form a directed knowledge graph where \verb|Subject| and \verb|Object| are nodes while \verb|Predicate| describes an edge. We use automatic reasoning on this graph to either infer new knowledge based on existing knowledge, or to verify the correctness of the knowledge base. Both avenues are possible thanks to the use of a meta-layer describing the graph (e.g., restrictions on concepts and relations), specified in a knowledge representation language, such as RDFS (Resource Description Framework Schema) or OWL (Web Ontology Language)~\cite{hogan2020knowledge}. The structure of the graph can be divided into two parts: \verb|TBox| (terminology: concepts and relations definition) and \verb|ABox| (assertions: statements about specific instances). In our aeroengine case, a statement \lstinline{Compressor isDepictedOn Plot} is part of the \verb|TBox|, while \lstinline{HPC_of_Engine_3 isDepictedOn Plot_4} belongs to the \verb|ABox|. The \verb|TBox| for the discussed system is shown in Fig.~\ref{fig:graph}.

We extract knowledge from the knowledge graph using the SQL-like language SPARQL (SPARQL Protocol and RDF Query Language)~\cite{sparql11-overview}. Listing~\ref{lst:sparql} presents a SPARQL query template for a \lstinline{intent:get_value} set of user commands (see Listing~\ref{lst:rasanlu}). We prepared such SPARQL query templates for all intent/command types.

An important task is to insert precise values into placeholders. Since these values are extracted by RASA NLU from automatically generated text from the speech signal, these values may be incorrectly recognized and therefore not match values in the knowledge graph. We solve this problem by replacing these values with the closest available values in the knowledge graph before constructing a query against the knowledge graph.

\begin{lstlisting}[caption=SPARQL query for getting a value requested by an user. Uppercase entities within brackets are placeholders for values gathered from the user's command., label=lst:sparql, float=htp]
SELECT ?ID ?subs ?val
WHERE { ?chara rdfs:label [CHARACTERISTIC] .
          ?subs  rdfs:label [SUBSYSTEM] .
          ?subs_inst a ?subs ;
                ?chara ?val ;
                aero:isPartOf ?engine .
          ?engine aero:VR_ID ?ID ;
                rdfs:label [ENGINE_NAME] .
\end{lstlisting}

\noindent The steps making up the speech bundle server flow are as follows:
\begin{enumerate}
    \item The server receives the command as text from the speech-to-text service.
    \item The command is processed by the RASA NLU model. Intent type and values for placeholders are returned.
    \item Values are cleared:
    (a) for numbers the text string is parsed to extract the numbers;
    (b) for enums, the closest value is selected from a hard-coded set;
    (c) for other text strings, the closest value is extracted from the knowledge graph.
    \item The final query is prepared using a template for the identified command type with placeholders replaced with cleared values. This query is then processed against the knowledge graph.
    \item The system formats the response for the user. This respoonse is closely linked to the values returned by the query (see Listing~\ref{lst:sparql}):
    \lstinline{engineID} is \lstinline{?ID},
    \lstinline{subsystem} is \lstinline{?subs},
    and \lstinline{message} is created using a template for a specific command type, involving, for example, \lstinline{?val}.
\end{enumerate}

The knowledge graph server not only allows the user to submit queries (using \lstinline{ask} as an endpoint). There are in addition three further JSON-RPC endpoints that can be accessed to manipulate the knowledge graph. In our current prototype, they are utilized to initialize the knowledge base. In the future, they can also be used to manipulate the knowledge graph  on-the-fly. By using these methods, the knowledge graph server can be treated as a black box, which can be fully operated by means of appropriate queries without any need for low-level manipulation of knowledge graph files or a codebase:
\begin{itemize}
    \item \lstinline{add_engine} provides the possibility to add information about new aeroengines (IDs, subsystems and their characteristics) to the knowledge graph. This endpoint is currently used to fill the knowledge graph when the system starts. However, it can also be used in a dynamic adaptation of the system to accommodate introducing new aeroengines on-the-fly to the Jarvis system.
    \item \lstinline{add_update_method} enables placing arbitrary Python code in the knowledge graph (see the ``Function'' node in Fig.~\ref{fig:graph}) that can be used to calculate specific characteristics with the use of other characteristics' values (see Listing~\ref{lst:add_update_method})
    \item \lstinline{update_values} allows changing specific values of the selected aeroengine during the system's operation. Other values are recalculated if necessary according to the defined update methods.
\end{itemize}

\begin{lstlisting}[caption=Sample JSON for the \lstinline{add_update_method} endpoint., label=lst:add_update_method, float=htp]
{
    "characteristic": "SS"     # Shaft Speed
    "func_args": ["MF", "PR"]  # Mass Flow, Pressure Ratio
    "func_code": "
    import some_library

    def func(sub, mf, pr):     # first value is always subsystem; then there are all func_args
        new_val = None
        if sub == 'LPC':
            new_val = # somehow calculate the value
        else:
            new_val = # calculate another way for other subsystems
        return new_val
    "
}
\end{lstlisting}

\section{DISCUSSION}
Our prior work~\cite{tadeja_ieee_2020} explored the deploying a VR environment for design and digital twinning in aeronautics.
This paper has presented an extended system, which we call Jarvis---a speech-enhanced VR demonstrator based on mining knowledge databases. In comparison to prior work~\cite{tadeja_ieee_2020} the system has been extended to incorporate a knowledge graph that supports voice queries from the user and which serves to amplify users abilities by allowing the use of voice interaction in conjunction with bimanual interaction.

As demonstrated in prior work \cite{tadeja_ieee_2020}, the gestures recognized by our system are sufficient to support the key tasks identified in the task analysis (see Subsection~\ref{tasks}): 1) visualize aeroengines; 2) analyze data; 3) explore performance data; 4) select aeroengine model; 5) inspect aeroengine model; and 6) manipulate aeroengine model.

While tasks 5) and 6) above are largely unaffected by the new voice interface, tasks 1)--4) are enhanced by voice interaction. For example, tasks 3) and 4) rely on the selection of appropriate data, for example, to select and take apart models with poor conditions in order to identify potential physical damage.

Such advanced features require the system to not only recognize the user's speech but to also infer further information based on the context in which the command was issued. This in turn is possible using the knowledge-based solution in the Jarvis system.  Interaction data and information on the current state of each of the compressors in each turboengine is captured in a knowledge-based system underpinned by a knowledge graph storing complex domain knowledge (see Subsection~\ref{sec:knowgraph}). This enables the Jarvis system to handle complex voice queries, such as ``what is the pressure ratio of the intermediate pressure compressor for Engine 5'' or ``calculate the average pressure ratio after 80 hours of flying time for HPC in Engine 3.'' The Jarvis system not only provides sufficient and precise answers to such queries but also allows a certain level of flexibility as the user can construct queries from a broader lexicon. This is possible because the context in which the user and the system are operating is concise and to some extent very strict, since the number of questions that  domain-expert users may ask in a given context is limited. In the future, the voice functionality could be extended with other, non-engineering commands, such as ``Recall/Hide the globe gizmo.''

A voice-operated system which proof-of-concept we are presenting here naturally has some limitations imposed by both hardware and software alike. These limitations can be categorized in three groups: (1) gaze- and eye-tracking; (2) gesture-tracking and gesture recognition; and (3) speech-to-text processing.

Regarding limitation 1), the Oculus Rift~\cite{oculus} VR headset does not provide built-in gaze-tracking or eye-tracking capabilities. Hence, we developed an approximation of a gaze-tracking feature by placing a cross-hair at a constant distance in front of the user's face (see Fig.~\ref{fig:full_view}). Using integrated eye-tracking hardware would most likely be more accurate and robust than our solution and consequently increase the level of fidelity of the Jarvis system. We anticipate this limitation will be alleviated with updated VR headsets in the near future.

Regarding limitation 2), among the gestures recognized by our system (see Fig.~\ref{fig:gestures}) the most frequently used gestures are the single and double pinch gesture~\cite{tadeja_ieee_2020}. Some studies suggest that replacing a pinch gesture with a grasp gesture may yield a better user experience in some circumstances \cite{jude_grasp_2016}. We leave such an investigation for the Jarvis system as future work.

Regarding limitation 3), speech-to-text processing occurs in discrete 60-second chunks instead of as a continuous process. The implementation is based on the conversation services provided by Google Cloud\footnote{\protect\url{https://cloud.google.com/speech-to-text}}. The free resources used by our system impose these limitations in length and size on the converted speech and this limitation can potentially be limited by licensing a bespoke solution.

\section{CONCLUSIONS}
In this paper we have presented the design and implementation of the Jarvis system---a state-of-the-art speech-enhanced VR demonstrator based on mining knowledge databases. This work represents one of the first attempts at aeroengine analysis in a bespoke voice-enabled VR environment and hopefully brings us a step closer to a real-life version of the J.A.R.V.I.S.~from the movies for aeroengine analytics.

The Jarvis system is equipped with a new bimanually operated globe gizmo (see Fig.~\ref{fig:globe_gizmo}) that allows the user to efficiently and swiftly select the  appropriate turboengine 3D CAD models~\cite{heyns} of interest. Further, this gizmo, if coupled with streamed GPS and other telemetry data, can provide the user with a broader view of the entire fleet of airplanes, including meta-information such as their current position and operational status.

We enhanced the Jarvis system with a robust and flexible subsystem that allows domain experts to issue voice commands while working within the VR environment. To achieve this goal, the system not only has to be able to recognize the user's speech. It also has to perform an analysis to gain an understanding of the information incorporated in the command and the situational context. This analysis allows the system to automatically infer further knowledge and act accordingly (see Subsection~\ref{voice}). This functionality is bestowed by the knowledge-based subsystem, which is in turn facilitated by a knowledge graph (see Fig.~\ref{fig:graph}) that stores complex domain knowledge. Thanks to this solution the Jarvis system is well-equipped to promptly answer a number of sensible context-aware queries.

\section*{ACKNOWLEDGEMENTS}
This work was supported by studentships from the Engineering and Physical Sciences Research Council (EPSRC-1788814), the Cambridge European \& Trinity Hall Scholarship, and the Cambridge Philosophical Society Research Studentship. In addition, the work of Yupu Lu was supported in part by the Tsinghua Academic Fund for Undergraduate Overseas Studies and the Fund from Tsien Excellence in Engineering Program. This work was supported by Wave 1 of The UKRI Strategic Priorities Fund under the EPSRC Grant EP/T001569/1, particularly the ``Digital Twinning in Aeronautics'' theme within that grant \& The Alan Turing Institute’.

The authors would also like to thank Kacper \L odzi\'{n}ski for his work on the Python code responsible for data exchange with the speech-to-text services offered by the Google Cloud API, and to Przemys\l aw Stachura for his help with the software debugging.

\bibliographystyle{asmems4}
\bibliography{references}

\end{document}